\documentclass[aps,prl,twocolumn,showpacs,amsmath,amssymb,amsfonts,nofootinbib,long,floatfix]{revtex4}
\usepackage{epsfig,latexsym,bm}

\newcommand\eV{\mbox{eV}}
\newcommand\GeV{\mbox{GeV}}

\newcommand\A{\mathbf{A}}
\newcommand\B{\mathbf{B}}
\newcommand\x{\mathbf{x}}

\newcommand\kk{\mathbf{k}}
\newcommand\ee{{\boldsymbol \varepsilon}}
\newcommand\mPl{m_{\rm Pl}}

\newcommand\E{\mathbf{E}}
\newcommand\D{\mathbf{D}}

\newcommand\HH{\mathbf{H}}

\newcommand\n{\mathbf{n}}

\begin{document}

\title{Electromagnetism with dimension-five operators}

\author{Leonardo Campanelli$^{1}$}
\email{leonardo.campanelli@ba.infn.it}
\affiliation{$^1$Dipartimento di Fisica, Universit\`{a} di Bari, I-70126 Bari, Italy}

\date{\today}

%***********************************   Abstract   ********************************************%

\begin{abstract}
We derive, in curved spacetime, the most general Lorentz-violating electromagnetic Lagrangian
containing dimension-five operators with one more derivative than the Maxwell term
in the hypothesis that Lorentz symmetry is broken by a background four-vector $n_\mu$.
We then study, for the case of isotropic $n_\mu$, the generation of cosmic magnetic fields at
inflation and cosmic birefringence.
In the limiting case of Minkowski spacetime, we find that other than the CPT-odd Myers-Pospelov term,
there exists another CPT-odd term that gives rise to nontrivial dispersion and constitutive relations.
\end{abstract}

%*********************************************************************************************%


%11.30.Cp -> Lorentz invariance
%98.80.-k -> Cosmology

\maketitle

%*********************************************************************************************%

\section{I. Introduction}

The search of Lorentz and CPT symmetry violation effects at low-energy scales
has received a renewed interest in recent years due to the improved
sensitivities of both terrestrial experiments and astrophysical observations~\cite{Data,Liberati}.
The motivation behind the investigation of such possible effects
is that in some theories
that aim to give a quantum-consistent description of gravity,
such as loop quantum gravity~\cite{Gambini} and string theory~\cite{Kostelecky3},
the breakdown of Lorentz and CPT symmetries is expected to take place
around the Planck scale $\mPl \simeq 10^{19} \GeV$.

At much lower energies, attainable by present-day experiments,
the breaking of such fundamental symmetries may manifest itself, for example,
by a modification of the standard dispersion relations of freely propagating particles
such as photons.
Irrespective of the underlying fundamental quantum gravity theory, however,
effects of Lorentz and CPT symmetry violation can be studied in the
framework of low-energy effective field theories~\cite{Potting,Kostelecky2}.

The simplest of such effective theories is the so-called minimal Standard-Model Extension (SME),
which represents the extension of the standard model of particle physics that incorporates
all possible Lorentz-violating renormalizable (up to dimension 4)
operators in Minkowski~\cite{Colladay} and curved~\cite{Kostelecky2} spacetimes.

A theory, known as Standard-Model Extension, with Lorentz-violating nonrenormalizable
(dimension 5 and more) operators in Minkowski spacetime has also been recently constructed
for the case of photons~\cite{Kostelecky1}
and fermions~\cite{Kostelecky4} and represents a generalization of the so-called Myers-Pospelov
model~\cite{MP} where only dimension-5 operators were considered.

In particular, the electromagnetic Myers-Pospelov Lagrangian was constructed
by relying upon the following six criteria~\cite{MP}:
($i$) quadratic in the electromagnetic field,
($ii$) one more derivative than the usual Maxwell term,
($iii$) gauge invariant,
($iv$) Lorentz invariant, except for the presence of an external four-vector $n_\mu$,
($v$) not reducible to lower dimension operators by the equations of motion, and
($vi$) not reducible to a total derivative.

The aim of this paper is to consider the extension
of the electromagnetic Myers-Pospelov model to the case of a general curved background spacetime.
This is useful when testing the theory in a cosmological context using, for example,
data of cosmic birefringence of the Cosmic Microwave Background (CMB) radiation (see, e.g.,~\cite{Gubitosi,Mewes}),
or when studying the creation of large-scale cosmic magnetic fields~\cite{ReviewCMF}
during the inflation era.

Interestingly enough, we will also find, when reducing ourselves to the Minkowski case, that other
than the usual Myers-Pospelov electromagnetic Lagrangian term there exists another CPT-odd term
satisfying the six criteria above enunciated and neglected in the seminal paper~\cite{MP}.
We will then analyze this term by finding the photon dispersion relations, fundamental for the
quantization of the theory,
and showing that, at low energies and for the case of isotropic $n_\mu$, the propagation of light
in vacuum is formally equal to the propagation of light in a particular bi-isotropic medium
known as ``Pasteur medium.''

\section{II. Curved spacetime}

In Minkowski spacetime, (particle) Lorentz violation is implemented
by coupling physical fields to constant spacetime tensors $n^{a_1a_2...a_n}$,
called external or background tensors.
The passage from Minkowski to a general curved spacetime is obtained
via the vierbein $e^{\mu}_{\;\,a}$,
\footnote{The vierbein satisfies the condition $e^{\mu}_{\;\,a} e_{b\mu} = \eta_{ab}$, and
is such that $g_{\mu\nu} = e_\mu^{\;\,a} \, e_\nu^{\;\,b} \, \eta_{ab}$,
where $\eta_{ab}$ and $g_{\mu\nu}$ are the metric tensors in Minkowski and general curved spacetimes, respectively.}
$n^{\mu_1\mu_2...\mu_n} = e^{\mu_1}_{\;\,a_1} e^{\mu_2}_{\;\,a_2} ... e^{\mu_n}_{\;\,a_n} n^{a_1a_2...a_n}$~\cite{Kostelecky2}.
(In this section, indices in Minkowski spacetime are indicated with the first letters of the Latin alphabet and run from 0 to 3.
Indices in curved spacetimes are indicated with Greek letters and run from 0 to 3.
Latin indices from the middle of the alphabet run from 1 to 3 and indicate spatial components
of a given tensor.)
It is important to stress that, while in Minkowski spacetime the external
tensors are constant---in curved spacetime they are not---since
the vierbein $e^{\mu}_{\;\,a}(x)$ is, generally, a function of the spacetime position $x$.

\subsection{IIa. Lagrangian and equations of motion}

Let us construct the most general Lagrangian for the photon field $A_\mu$, ${\mathcal L}_{\rm em}^{(5)}$, containing
all gauge-invariant, Lorentz-violating terms,
which are quadratic in the electromagnetic field strength tensor
$F_{\mu \nu} = \partial_\mu A_\nu - \partial_\nu A_\mu$ and have just one more
derivative than the Maxwell term.
These terms are dimension-five operators that can generally be written as
\begin{equation}
\label{HD2}
{\mathcal L}_{\rm em}^{(5)} = \frac{1}{4\mPl} \, n^{\mu\nu\alpha\beta\gamma} F_{\mu \nu} F_{\alpha \beta ; \gamma},
\end{equation}
where a semicolon denotes covariant differentiation with respect to spacetime coordinates,
and $n^{\mu\nu\alpha\beta\gamma}$ is a dimensionless rank-5 background tensor
that breaks Lorentz symmetry. The presence of the Planck mass indicates
that we are assuming that Lorentz symmetry breaking
occurs at the Planck scale.
The external tensor $n^{\mu\nu\alpha\beta\gamma}$ is antisymmetric in the first two and second two indices.
Also, it can be taken to be antisymmetric for the interchange of the first two indices
with the second two indices. This is because the corresponding symmetric part would give,
after integrating by part the action $S_{\rm em} = \int \! d^4\!x \, \sqrt{-g} \, {\mathcal L}_{\rm em}$,
where $g$ is the determinant of the metric tensor and
${\mathcal L}_{\rm em} = -\frac{1}{4} F_{\mu \nu} F^{\mu \nu} + {\mathcal L}_{\rm em}^{(5)}$,
to a surface term plus a term proportional to
$n^{\mu\nu\alpha\beta\gamma}_{~~~~~~~ ;\gamma}  \, F_{\mu \nu} F_{\alpha \beta}$ which does
not contain, as required, an extra derivative of the Maxwell term.

Variation of the action gives the equation of motion
\begin{equation}
\label{DDD}
\mathcal{D}^{\mu \nu}_{\;\;\;\; ;\mu} = 0,
\end{equation}
where $\mathcal{D}^{\mu \nu} = F^{\mu \nu} - \mathcal{M}^{\mu \nu}$ is
the so-called ``displacement tensor,'' and
$\mathcal{M}^{\mu \nu} = n^{\mu\nu\alpha\beta\gamma} F_{\alpha \beta ;\gamma}
+ \frac12 \, F_{\alpha \beta} n^{\mu\nu\alpha\beta\gamma}_{~~~~~~~ ;\gamma}$
is the ``polarization-magnetization tensor.''
The Bianchi identities are $\widetilde{F}^{\mu\nu}_{~~~;\mu} = 0$,
where $\widetilde{F}^{\mu\nu} = \frac12 E^{\mu\nu\alpha\beta} F_{\alpha\beta}$
is the dual electromagnetic field strength tensor,
with $E^{\mu \nu \alpha \beta} = \varepsilon^{\mu \nu \alpha \beta}/\sqrt{-g}$ being
the totally antisymmetric tensor in four dimensions
and $\varepsilon^{\mu\nu\alpha\beta}$ the Levi-Civita
symbol (with $\varepsilon^{0123} = +1$).

Let us now assume that Lorentz symmetry is broken just by the presence of
a background four-vector $n_\mu$. By inspection, the tensor $n^{\mu\nu\alpha\beta\gamma}$
with the above discussed symmetry properties that can be constructed using $n_\mu$, the metric tensor
$g_{\mu\nu}$, and $E^{\mu\nu\alpha\beta}$ can be written as
\begin{eqnarray}
\label{HD3}
n^{\mu\nu\alpha\beta\gamma}
\!\!& = \!\!& \zeta \! \left( g^{\alpha[\mu}E^{\nu]\gamma\delta\beta}
      - g^{\beta[\mu}E^{\nu]\gamma\delta\alpha} \right) \! n_\delta \nonumber \\
\!\!& + \!\!& \xi \!\left( E^{\mu\nu\delta[\alpha}n^{\beta]}
      - E^{\alpha\beta\delta[\mu}n^{\nu]} \right) \! n_\delta n^\gamma,
\end{eqnarray}
where $\zeta$ and $\xi$ are dimensionless coupling constants
(which, without loss of generality, we assume to be positive definite),
and square brackets indicate antisymmetrization
of the indices enclosed, e.g.,
$T_{\mu_1 ... [\mu_i \mu_j] ... \mu_n} =
\frac12 (T_{\mu_1 ...\mu_i \mu_j... \mu_n} - T_{\mu_1 ... \mu_j \mu_i ... \mu_n})$.
Taking into account Eq.~(\ref{HD3}), we can recast Lagrangian~(\ref{HD2}) in the form
\begin{eqnarray}
\label{extra}
\!\!\!\!\!\!\!\!\! {\mathcal L}_{\rm em}^{(5)} \!\!& = &\!\!
\frac{\zeta}{2\mPl} \, \widetilde{n}^{\mu\alpha}_{~~~\beta} F_{\mu \nu} D_\alpha F^{\beta \nu} \nonumber \\
\!\!\!\!\!\!\!\!\! \!\!& + &\!\! \frac{\xi}{2\mPl} \, n^\mu n^\alpha n_\beta
(F_{\mu \nu} D_\alpha \widetilde{F}^{\beta \nu} - \widetilde{F}_{\mu \nu} D_\alpha F^{\beta \nu}),
\end{eqnarray}
where $\widetilde{n}^{\mu\alpha\beta} = n_\gamma \varepsilon^{\gamma\mu\alpha\beta}$
is the dual tensor of the vector $n^\mu$.
The term proportional to $\xi$ can be viewed as the general-covariant
generalization of the electromagnetic Myers-Pospelov Lagrangian (see Sec. IIIa).

Let us now restrict our analysis to the case of a spatially flat, Friedmann-Robertson-Walker universe,
described by the line element $ds^2 = a^2(d\eta^2 - d \x^2)$,
where $\eta$ is the conformal time and
$a(\eta)$ is the expansion parameter.
Since $g_{\mu\nu} = a^2 \eta_{\mu\nu}$, we can take for the vierbein
$e_\mu^{\;\,b} = a\delta_\mu^b$, %$e^\mu_{\;\,c} = \delta^\mu_c /a$,
where $\delta_\mu^b$ is the Kronecker delta.
Also, we consider only the case of timelike external vector $n_b$,
$n_b = (1,0,0,0)$ (to be close to the original analysis in~\cite{MP}),
so that $n_\mu = e_{\mu}^{\;\,b} \, n_b = (a,0,0,0)$ and $n^{\mu} = (1/a,0,0,0)$.
%$n^{\mu} = g^{\mu\nu} n_\nu = (1/a,0,0,0)$.

Working in Coulomb gauge, $A_\mu = (0,\A)$ with $\partial_i A_i = 0$,
the equation of motion~(\ref{DDD}) becomes
\begin{equation}
\label{vectorAeq}
\A'' - \nabla^2 \A + \mathfrak{D}[\nabla \times \A] = 0,
\end{equation}
where a prime denotes differentiation with respect to the conformal time, $\nabla$ is the nabla operator with respect
to comoving coordinates, and
$\mathfrak{D}$ is the second order differential operator
\begin{eqnarray}
\label{operator}
\mathfrak{D} \!\!& = &\!\! (\tilde{g}_1 + 2 \tilde{g}_2) \frac{\partial^2}{\partial \eta^2}
- \tilde{g}_1 \nabla^2
-  (\tilde{g}_1 + 2 \tilde{g}_2) \mathcal{H} \frac{\partial}{\partial \eta} \nonumber \\
\!\!& + &\!\! \left(\! 2\tilde{g}_1 + \frac12 \tilde{g}_2 \! \right) \! (\mathcal{H}^2 - \mathcal{H}').
\end{eqnarray}
Here, $\tilde{g}_i = g_i/a$, $g_1 = \zeta/\mPl$, $g_2 = \xi/\mPl$, and $\mathcal{H} = a'/a$.
In obtaining Eq.~(\ref{vectorAeq}), we used the fact that %$g_i' = -\mathcal{H} g_i$,
%and that
the nonzero Christoffel symbols are
$\Gamma^{i}_{j0} = \Gamma^{i}_{0j} = \Gamma^{0}_{ij} = \Gamma^{0}_{ji} = \delta_{ij} \Gamma^{0}_{00} = \delta_{ij} \mathcal{H}$,
so that the only nonzero components of $n_{\mu;\nu}$ are $n_{i;j} = - a \mathcal{H} \delta_{ij}$.

To solve Eq.~(\ref{vectorAeq}), let us expand the vector potential as
\begin{equation}
\label{A2}
{\A}(\eta,\x) = \sum_{\lambda=1}^2 \int \!\! \frac{d^3k}{(2\pi)^3 \sqrt{2|\kk|}} \, \ee_{\kk,\lambda} \,
A_{\lambda}(\eta,|\kk|) \, e^{i\kk \x} + \mbox{c.c.} ,
\end{equation}
where $\kk$ is the comoving wave number, and $\ee_{\kk,\lambda}$ are the standard circular polarization vectors.
\footnote{The circular polarization vectors are defined through the relations
$\kk \cdot \ee_{\kk,\lambda} = 0$,
$\ee_{\kk,\lambda} \cdot \ee_{\kk,\lambda'}^* = \delta_{\lambda \lambda'}$, and
$\sum_{\lambda=1}^2 (\ee_{\kk,\lambda})_i (\ee_{\kk,\lambda'}^*)_j = \delta_{ij} - \hat{k}_i \hat{k}_j$,
and satisfy the useful properties
$\ee_{-\kk,\lambda}^* = -\ee_{\kk,\lambda}$ and
$i \hat{\kk} \times \ee_{\kk,\lambda} = (-1)^{\lambda + 1} \ee_{\kk,\lambda}$.}
Inserting in Eq.~(\ref{vectorAeq}), we get the equation of motion for the
two photon polarization states $A_\lambda$,
\begin{equation}
\label{new1}
(1-\beta_3)A''_\lambda -\beta'_3 A'_\lambda + [(1-\beta_1) |\kk|^2 -\beta''_2\,]  A_\lambda = 0,
\end{equation}
where
$\beta_1 = (-1)^{\lambda} \, \tilde{g}_1 |\kk|$,
$\beta_2 = (-1)^{\lambda}(2\tilde{g}_1 + \tilde{g}_2/2) |\kk|$, and
$\beta_3 = (-1)^{\lambda} (\tilde{g}_1 + 2\tilde{g}_2) |\kk|$.

It is useful, for the following discussion, to
define the rescaled electromagnetic field
$\psi_\lambda = \sqrt{1-\beta_3} A_\lambda$.
Inserting in Eq.~(\ref{new1}), we find that it satisfies the equation of motion
\begin{equation}
\label{psi2}
\psi''_\lambda = U_\lambda \psi_\lambda,
\end{equation}
where
\begin{equation}
\label{new2}
U_\lambda = -\frac{1-\beta_1}{1-\beta_3} \, |\kk|^2 + \frac{\beta''_2}{1-\beta_3}
+ \frac{1}{\sqrt{1-\beta_3}} \frac{\partial^2}{\partial \eta^2} \sqrt{1-\beta_3} \, .
\end{equation}
Equation~(\ref{psi2}) is formally equal to the one-dimensional Schrodinger equation
with zero energy and potential energy $U_\lambda$,
$\eta$ taking the place of the spatial coordinate, and $g_i$, $|\kk|$, and $\lambda$ playing the role of
free constant parameters.

\subsection{IIb. Cosmic magnetic fields}

Microgauss magnetic fields are observed in all large-scale gravitationally bound
systems, such as galaxies and clusters of galaxies, and there are hints that they exist
even in cosmic voids~\cite{ReviewCMF}. Their large correlation, up to Mpc scales, and
their ubiquity supports the idea that they have been created in the early
universe~\cite{Turner-Widrow}, presumably during inflation~\cite{Banerjee}.

It is generally believed that a necessary condition to generate inflationary cosmic magnetic fields is
to introduce new terms in the photon Lagrangian that break the conformal invariance
of standard electromagnetism~\cite{Turner-Widrow}
(see~\cite{Dolgov,Barrow,Campanelli} for mechanisms that
work without resorting to nonstandard physics ). This may allow for a ``superadiabatic''
amplification of large-scale magnetic fluctuation during inflation that then may eventually
evolve and survive from the end of inflation until today~\cite{Kahniashvili}.

The Lorentz-violating terms in Eq.~(\ref{extra}) naturally break electromagnetic conformal invariance,
so it is interesting to see if they may also be responsible for the creation of cosmic magnetic fields
(see~\cite{Campanelli2} for other Lorentz-violating electromagnetic Lagrangians that can give
rise to large-scale magnetic fields).

Let us consider, for the sake of simplicity, the case of de Sitter inflation.
In this phase, the conformal time is inversely proportional to the expansion parameter,
$\eta = - 1/aH_{\rm dS}$, while the Hubble parameter $H_{\rm dS}$ is a constant.
On large super-Hubble scales, $-|\kk|\eta \ll 1$, which are the only important scales
for cosmic magnetic fields~\cite{Turner-Widrow}, the ``potential energy'' $U_\lambda$ reads
$U_\lambda = -|\kk|^2 \! \left[ 1+ b^2 + \mathcal{O}(|\kk|\eta) \right]$,
where $b = \frac12 (-1)^\lambda (\zeta + 2\xi) H_{\rm dS}/\mPl$. In this case,
the solution of Eq.~(\ref{psi2}) is
\begin{equation}
\label{CMF2}
A_\lambda = c(|\kk|,b) \, e^{-i |\kk|\eta \, \sqrt{1-b^2}},
\end{equation}
where the coefficient $c(|\kk|,b)$ is a normalization constant
that must reduce to unity for vanishing coupling constants,
$\lim_{b \rightarrow 0} c(|\kk|,b) = 1$, so to have the usual plane-wave solution
$A_\lambda = e^{-i |\kk|\eta}$ in the limiting case of Maxwell theory~\cite{Birrell-Davies}.

Consideration of graviton production requires that the scale of inflation $M$
is below $10^{16}\GeV$~\cite{Planck,BICEP2}.
Assuming instantaneous reheating, so that
$M^4 = \rho_{\rm inf} = 3H_{\rm dS}^2/8\pi G$, where $\rho_{\rm inf}$ is the
total energy density during inflation and $G = 1/\mPl^2$ the Newton's constant,
we have $b \lesssim 10^{-6} (\zeta + 2\xi)$. %$H_{\rm dS}/\mPl \lesssim 10^{-6}$.
Accordingly and even in the case of strong coupling, $\zeta \sim \xi \sim 1$, only small values of
$b$ are allowed giving $|A_\lambda|^2 \sim 1$.

We conclude that the inflation-produced magnetic field evolves adiabatically
on super-Hubble scales---superadiabatic amplification occurs when
$|A_\lambda|^2 \propto f(a)$ with $f$ an increasing function of $a$---and that
the production of cosmic magnetic fields is inhibited in the model at hand.

\subsection{IIc. Cosmic birefringence}

Lorentz-violating electromagnetic theories can give rise to
effects of cosmic birefringence (see, e.g.,~\cite{Mewes2}): the left- and right-circular polarized components of light,
$A_\lambda$, propagate differently {\it in vacuo}, resulting in potentially observable effects,
such as the rotation of the polarization plane of (partially) linearly polarized light,
such as the CMB radiation.

At low momenta compared to the Planck scale,
the potential energy %in Eq.~(\ref{new2})
reads
$U_\lambda = -|\kk|^2 -
2(-1)^\lambda \tilde{g}_2|\kk|^3 \left[ 1 + \mathcal{O}(\mathcal{H}^2/|\kk|^2) + \mathcal{O}(\mathcal{H}'/|\kk|^2) \right]$
to the lowest order in $|\kk|/\mPl$, with the terms of
order $\mathcal{H}^2/|\kk|^2$ and $\mathcal{H}'/|\kk|^2$
coming from the last two terms in Eq.~(\ref{new2}).
For typical photons of CMB radiation, these terms are vanishingly small when compared to the first term in Eq.~(\ref{new2}).
To see this, let us introduce the redshift $z = 1/(1+a)$, so that
$\mathcal{H}(z) = H_0 (1+z)^{-1} E(z)$ and
$\mathcal{H}'(z) = H_0^2 (1+z)^{-2} \left[1 - \frac12 (1+z) \partial_z \right] \! E^2(z)$,
where $E(z) = H(z)/H_0$ is the Hubble parameter $H(z)$ normalized to the Hubble constant $H_0$.
In a spatially flat universe dominated by dark matter and cosmological constant,
we have $E(z) = \sqrt{\Omega_m (1+z)^3 + \Omega_\Lambda}$, where $\Omega_m$ and $\Omega_\Lambda$
are the usual energy density parameters.
\footnote{We recall that $\Omega_m = \rho_m^{(0)}/\rho_{\rm cr}^{(0)}$ and
$\Omega_\Lambda = \rho_\Lambda/\rho_{\rm cr}^{(0)}$, where $\rho_m^{(0)}$ is the present value of the
energy density of matter, $\rho_\Lambda = \Lambda/8\pi G$, with $\Lambda$ being
the cosmological constant, and $\rho_{\rm cr}^{(0)} = 3H_0^2/8\pi G$ is the actual critical energy density.
Moreover, $\Omega_m + \Omega_\Lambda = 1$ in a spatially flat Friedmann-Robertson-Walker universe.}
Accordingly, we get $\mathcal{H}^2(z) = H_0^2 [\Omega_m (1+z) + \Omega_\Lambda]$ and
$\mathcal{H}'(z) = H_0^2 [-\frac12 \Omega_m (1+z) + \Omega_\Lambda]$.
Taking into account that $H_0 \sim 10^{-42}\GeV$ and that $\Omega_m$ and $\Omega_\Lambda$ are order one
parameters, we obtain that $\mathcal{H}^2/|\kk|^2$ and $\mathcal{H}'/|\kk|^2$ are at most of order
$z_{\rm dec} H_0^2/\omega_{\rm CMB}^2 \sim 10^{-55}$,
where we used the fact that $|\kk| \sim \omega_{\rm CMB}$, with $\omega_{\rm CMB} \sim 10^{-4} \eV$ being the
typical energy of CMB photons and $z_{\rm dec} \sim 10^3$ their (maximum) redshift at decoupling~\cite{Kolb-Turner}.

In order to find how the two polarized photon states evolve in time,
we need the solution of the equation of motion~(\ref{psi2}). An analytical expression of
such a solution cannot be found in a universe dominated by dark matter and cosmological constant.
Nevertheless and for our purpose, it suffices to solve it in Wentzel-Kramers-Brillouin (WKB) (semiclassical) 
approximation~\cite{Landau}. We find
\begin{equation}
\label{psiWKB}
\psi_\lambda^{(\rm WKB)} = c_\lambda(|\kk|,\tilde{g}_i) \, e^{-i \!\int^\eta \! d\eta' \sqrt{-U_\lambda(\eta')}},
\end{equation}
where $c_\lambda(|\kk|,\tilde{g}_i)$ is a real normalization constant whose explicit expression is inessential for the
following discussion.
The above solution is valid whenever $|U'_\lambda/2U_\lambda^{3/2}| \ll 1$~\cite{Landau},
a condition that is certainly satisfied in our case since $|U'_\lambda/2U_\lambda^{3/2}| \sim \xi H(z)/\mPl \ll 1$.

Now, the rotation angle of the polarization plane of linearly polarized light,
$\Delta \alpha$, is defined as the phase difference between left-handed and right-handed
photons emitted from a source at the time $\eta_e$ and observed at the time $\eta_o$.
From the above WKB solution, we then get
\begin{eqnarray}
\label{g5}
\Delta \alpha & = & \int_{\eta_e}^{\eta_o} \! d\eta \! \left[ \! \sqrt{-U_2(\eta)} - \sqrt{-U_1(\eta)} \, \right] \nonumber \\
& = & \frac{2\xi|\kk|^2}{\mPl H_0} \int^z_0 \! dz' \, \frac{1+z'}{E(z')},
\end{eqnarray}
where $z = z_e$ is the redshift of the source ($z_{\rm dec}$ for CMB photons) which emits photons
reaching the observer today at redshift $z_o = 0$. Interestingly enough, the rotation angle depends
only on $\xi$. This means that only the Myers-Pospelov Lagrangian term in Eq.~(\ref{extra})
contributes, to the leading order in $|\kk|/\mPl$, to the effect of cosmic birefringence.
Equation~(\ref{g5}) is in agreement with the result quoted in the
literature for the Myers-Pospelov case (see, e.g.,~\cite{Gubitosi})
and obtained, with the aid of an appropriate redshift-dependent scaling of the
photon momentum, by extending the exact result in Minkowski spacetime~\cite{Kostelecky1}
to the Friedmann-Robertson-Walker case.

\section{III. Minkowski spacetime}

Let us now specialize some of the above results to the case of Minkowski spacetime.
In particular, we are interested in the dispersion relations of freely propagating photons,
since they are an essential ingredient for the construction of a self-consistent
Lorenz-violating, quantum electrodynamic theory~\cite{Reyes}.

\subsection{IIIa. Lagrangian and equations of motion}

In Minkowski spacetime, the coefficients for Lorentz violation
are constants, $\partial_\gamma n^{\mu\nu\alpha\beta\gamma} = 0$.
(In this section, indices in Minkowski spacetime are indicated with Greek letters and run from 0 to 3.
Latin indices from the middle of the alphabet run from 1 to 3 and indicate spatial components
of a given tensor.)
Taking into account Eq.~(\ref{HD3}) we can then recast Lagrangian~(\ref{HD2}) in the form
\begin{equation}
\label{HDX}
{\mathcal L}_{\rm em}^{(5)} = \frac{\zeta}{2\mPl} \, \widetilde{n}^{\mu\alpha}_{~~~\beta} F_{\mu \nu} \partial_\alpha F^{\beta \nu}
+ \frac{\xi}{\mPl} \, n^\mu  F_{\mu \nu} n^\alpha \partial_\alpha n_\beta \widetilde{F}^{\beta \nu}.
\end{equation}
Both terms in Eq.~(\ref{HDX}) are dimension-five, CPT-odd operators and satisfy the
six Myers-Pospelov criteria (see the Introduction).
In~\cite{MP}, however, just the second term  was considered,
and it is now known as the Myers-Pospelov Lagrangian.
Interestingly enough, it has been shown in~\cite{Mariz} (see also~\cite{Mariz2}) that the first term in
Lagrangian~(\ref{HDX}) can be induced through radiative corrections from the fermion sector of the minimal SME.
Therefore, that term cannot be generally neglected
when constructing a low-energy, Lorenz-violating effective field theory with dimension-five operators.

Before proceeding further, let us frame our result~(\ref{HDX}) in the context of the SME.
To this end, we rewrite Lagrangian~(\ref{HD2}), in Minkowski spacetime, as
${\mathcal L}_{\rm em}^{(5)} =
\frac{1}{2\mPl} A_\lambda \, \overset{\circ}{n} \, \! ^{\alpha\lambda\mu\nu\beta} \partial_\alpha \partial_\beta F_{\mu \nu}$,
where the external tensor $\overset{\circ}{n} \, \!  ^{\mu\nu\alpha\beta\gamma}$ is antisymmetric in the second two indices
and symmetric for the interchange of the first with the last index.
When Lorentz symmetry is broken by the four-vector $n_\mu$, we have
$\overset{\circ}{n} \, \!  ^{\mu\nu\alpha\beta\gamma} = \zeta \, \epsilon^{\gamma\}\delta\nu[\alpha}\eta^{\beta]\{\mu}n_\delta -
\xi \, \epsilon^{\alpha\beta\delta\nu}n^{\mu}n^{\gamma}n_{\delta}$,
where $\zeta$ and $\xi$ coincide with the coupling constants in Eq.~(\ref{HDX}), and square brackets $\{...\}$ indicate symmetrization
of the indices enclosed, e.g., $T_{\mu_1 ... \{\mu_i \mu_j\} ... \mu_n} =
\frac12 (T_{\mu_1 ...\mu_i \mu_j... \mu_n} + T_{\mu_1 ... \mu_j \mu_i ... \mu_n})$ and
$T_{\mu_1\} ... \{\mu_n} =
\frac12 (T_{\mu_1 ... \mu_n} + T_{\mu_n ... \mu_1})$.
In the SME notations of~\cite{Kostelecky1},
Lagrangian~(\ref{HDX}) is then obtained by
taking the nonzero coefficients for Lorentz violation to be
$(k_{AF}^{(5)})^{\kappa\mu\nu} =
\frac{1}{\mPl} \!\left( \widetilde{n}^{\kappa \mu\nu} - \frac15 \, \eta^{\kappa\{\mu} \widetilde{n}_{\alpha} \! \, ^{\nu \} \alpha}
- \frac15 \, \widetilde{n}_{\alpha} \! \, ^{ \alpha \{\nu} \eta^{\mu \} \kappa} \right)
= -\frac{\zeta}{3\mPl} \!\left( n^\kappa \eta^{\mu\nu} - \frac25 \eta^{\kappa\{\mu}n^{\nu\}} \! \right)
- \frac{\xi}{\mPl} \!\left( n^\kappa n^{\mu} n^{\nu} \! - \frac25 n^2 \eta^{\kappa\{\mu}n^{\nu\}} \! \right) \! ,$
where we have defined
$\widetilde{n}_{\kappa}^{~~\mu\nu} = \frac{1}{3!} \, \varepsilon_{\kappa\alpha\beta\gamma} \,
\overset{\circ}{n} \, \! ^{\mu\alpha\beta\gamma\nu}$.

The equations of motion~(\ref{DDD}) can be explicitly written in Minkowski spacetime as
\begin{equation}
\label{mot} \partial^\mu F_{\mu\nu} = \frac{\zeta}{\mPl} \, \widetilde{n}_{\nu\alpha\gamma} \partial^\gamma \partial_\beta F^{\alpha\beta}
+ \frac{\xi}{\mPl} \, \widetilde{n}_{\nu\alpha\beta} (n \cdot \partial)^2 F^{\alpha\beta},
\end{equation}
and then conveniently rewritten in three-dimensional vectorial form as
\begin{eqnarray}
\label{g1}
\!\!\!\!\!\!\!\!\!\!\!\!\!\!\!
&& \nabla \cdot \E = - \left[\,g_1 \square + 2g_2 (n \cdot \partial)^2\right] \!(\n \cdot \B), \\
\label{g2}
\!\!\!\!\!\!\!\!\!\!\!\!\!\!\!
&& \dot{\E} -\nabla \times \B = \left[\,g_1 \square + 2g_2 (n \cdot \partial)^2\right] \!(n^0 \B - \n \times \E).
\end{eqnarray}
Here, $\E = (E_1,E_2,E_3)$ and $\B = (B_1,B_2,B_3)$, where $E_i = -F_{0i}$ and
$B_i = \frac12 \, \epsilon_{ijk} F_{jk}$ are the electric and magnetic fields,
and a dot indicates the derivative with respect to time.
Equations~(\ref{g1}) and (\ref{g2}) are the $\nu = 0$ and $\nu = i$ components of Eq.~(\ref{mot}),
respectively. The Bianchi identities are $\nabla \cdot \B = 0$ and $\dot{\B} = -\nabla \times \E$.

\subsection{IIIb. Dispersion relations}

Taking the time derivative of Eq.~(\ref{g2}) and using the Bianchi identities,
we get
%$\square \E - \nabla (\nabla \cdot \E) +
%\left[\,g_1 \square + 2g_2 (n \cdot \partial)^2 \right] \!(n^0 \nabla \times \E - \n \times \dot{\E}) = 0$,
%
\begin{equation}
\label{mot2} \square \E - \nabla (\nabla \cdot \E) +
\left[\,g_1 \square + 2g_2 (n \cdot \partial)^2 \right] \!(n^0 \nabla \times \E - \n \times \dot{\E}) = 0,
\end{equation}
where $n^\mu = (n^0, \n)$.
To find the photon dispersion relations, we work in momentum space
by considering the ansatz $\E(x) = \E(k) \, e^{-ikx}$. Inserting in the above equation, %Eq.~(\ref{mot2}),
we get $M_{ij} E_j = 0$, where
the Hermitian matrix $M_{ij}$ is given by
$M_{ij} =  k^2 \delta_{ij} + k_i k_j + i \varepsilon_{ijl} \! \left[\,g_1 k^2 + g_2 (n \cdot k)^2 \right] \!(n_0 k_l - \omega n_l)$.
The covariant dispersion relations come from the condition $\det M_{ij} = 0$~\cite{Colladay} and read
\begin{equation}
\label{dispersion}
(k^2)^2 - \left[\,g_1 k^2 + g_2 (n \cdot k)^2 \right]^2 \! \left[(n \cdot k)^2 - n^2 k^2 \right] = 0.
\end{equation}
Writing $k^\mu = (\omega, \kk)$ and solving for $\omega$, one can obtain the frequency solutions.
The case $g_1 = 0$ has been fully analyzed in~\cite{Reyes}.
There, it has been shown that for a timelike $n_\mu$ the theory is unstable and violates causality,
while for a lightlike $n_\mu$ it contains extra
degrees of freedom that are nonanalytic in the perturbative parameter $g_2$,
leading to a nonunitary evolution at the quantum level.
Nonunitarity problems are expected to appear, however, only at high energies since,
as pointed out in~\cite{Kostelecky1}, extra (nonanalytic) modes play
a role only at Planck-scale energies.
The spacelike case, instead, is free from the above problems.

For the case $g_2 = 0$, we obtain $\omega = |\kk|$ for $\n = 0$, and
\begin{equation}
\label{solution}
\omega_{\pm} = \frac{1}{\n^2} \! \left( n^0 \kk \cdot \n \pm \sqrt{\frac{\n^2}{g_1^2} - n^2 |\kk \times \n|^2} \: \right)
\end{equation}
for $\n \neq 0$. In the latter case, there are 2 degrees of freedom that,
according to the ``Reyes analyticity criterion''~\cite{Reyes}, are nonanalytic in the perturbative parameter $g_1$.
(A frequency solution depending on $g_i$ is nonanalytic when it goes to infinity for $g_i \rightarrow 0$.)
Accordingly, as in the case $g_1 = 0$ and lightlike $n_\mu$,
we expect that the quantum theory develops problems of nonunitarity.

The case with both $g_1$ and $g_2$ different from zero cannot be handled analytically
in the general case. Nevertheless, analytical frequency solutions can be found, separately, for
the three specific cases of timelike, lightlike, and spacelike unit vector $n^\mu$.

($i$) Timelike case ($n^2 = 1$). For the sake of simplicity, let us consider the ``isotropic'' case $\n^2 = 0$.
\footnote{In this case,
the coefficients $g_1$ and $g_2$ are related to the ``isotropic coefficients'' $(\overset{\circ}{k}_{AF}^{(5)})_0$
and $(\overset{\circ}{k}_{AF}^{(5)})_2$ in the SME Lagrangian~\cite{Kostelecky1} by
$(\overset{\circ}{k}_{AF}^{(5)})_0 = \frac{\sqrt{4\pi}}{5} \, (g_1 + 3g_2)$ and
$(\overset{\circ}{k}_{AF}^{(5)})_2 = -\frac{\sqrt{4\pi}}{3} \, g_1$.
}
Solving Eq.~(\ref{dispersion}) for $\omega$, we obtain the frequency solutions
\begin{equation}
\label{AAA1}
\omega_\pm = |\kk|\sqrt{\frac{1 \pm g_1 |\kk|}{1 \pm (g_1+2g_2)|\kk|}} \, ,
\end{equation}
which correspond to right-handed and left-handed photons, respectively. To see this, we write
${\E}_\pm(t,\x) = \ee_{\kk,\pm} \, e^{-i\omega t + i\kk \x}$,
where $\ee_{\kk,\pm}$ are the circular polarization vectors introduced in the previous section
and $\pm$ stand for $\lambda = 1,2$.
Inserting in Eq.~(\ref{mot2}), we get
$\left[\, \omega^2 - |\kk|^2 \pm (g_1 + 2g_2) |\kk| \omega^2 \mp g_1 |\kk|^3 \right] \! \ee_{\kk,\pm} = 0$,
which is identically satisfied when $\omega$ takes on the values~(\ref{AAA1}).

For $1/(g_1 + 2g_2) < |\kk| < 1/g_1$, the frequency solution $\omega_{-}$ develops an imaginary part that,
as in the case $g_1 = 0$ discussed in~\cite{Reyes},
leads to the loss of unitarity and to instabilities.
Moreover, as it is straightforward to check, the group velocity
$v_{g,\pm} = d\omega_{\pm}/d|\kk| = (\omega_{\pm}/|\kk|)[1 \pm 2g_1|\kk| + (\omega_{\pm}/|\kk|)^2]/2(1 \pm g_1|\kk|)$
can exceed the speed of the light in the case of left-handed photons
introducing, then, problems of causality.
(The same situation happens for the case $g_1 = 0$~\cite{Reyes}.)

($ii$) Lightlike case ($n^2 = 0$). Let us take, for the sake of simplicity, $n^0 = 1$.
Since the expression of the six analytical solutions of the sixth order equation~(\ref{dispersion}) are cumbersome,
we write down only their asymptotic expressions in the limit of small momenta,
\begin{eqnarray}
\label{AAA5}
&& \!\!\!\!\!\!\!\!\!\!\!\!\!
\omega_{1,2} = |\kk| \! \left[ 1 \pm 8g_2 |\kk| \sin^6(\theta/2)  + \mathcal{O}(|\kk|^2/\mPl^2) \right] \!, \\
&& \!\!\!\!\!\!\!\!\!\!\!\!\!
\omega_{3,4} = -|\kk| \! \left[ 1 \pm 8g_2 |\kk|\cos^6(\theta/2)  + \mathcal{O}(|\kk|^2/\mPl^2) \right] \!, \\
&& \!\!\!\!\!\!\!\!\!\!\!\!\!
\omega_{5,6} = \pm \frac{1}{g_1 + 2g_2} \left[1 + \mathcal{O}(|\kk|^2/\mPl^2)\right] \!,
\end{eqnarray}
where $\theta$ is the angle between $\kk$ and $\n$.
As in the case $g_1 = 0$~\cite{Reyes}, there are well-behaved solutions ($\omega_{1,2}$ and $\omega_{3,4}$)
that approach the standard solutions ($\omega = \pm |\kk|$) in the limit $g_i \rightarrow 0$,
and extra solutions ($\omega_{5,6}$) that are nonanalytic in the perturbative parameters $g_i$.
The latter may introduce in the theory problems of nonunitarity at the quantum level.

($iii$) Spacelike case ($n^2 = -1$). For the sake of simplicity, let us consider the case $n^0 = 0$.
Also in this case, the expression of the frequency solutions are cumbersome.
Their asymptotic expressions are
\begin{eqnarray}
\label{AAA3}
&& \omega_{1,2} = |\kk| \! \left[1 \pm g_2 |\kk| |\cos\theta|^3 + \mathcal{O}(|\kk|^3/\mPl^3)\right]\!, \\
&& \omega_3 = \frac{1}{g_1} \left[ 1 + \mathcal{O}(|\kk|^2/\mPl^2)\right] \!.
\end{eqnarray}
In the Myers-Pospelov case ($g_1 = 0$)~\cite{Reyes} only the solutions $\omega_{1,2}$ are present. They are
well-behaved solutions since are analytic in $g_1$. In our case, instead, the first term in
Lagrangian~(\ref{HDX}) introduces an extra nonanalytic solution ($\omega_{3}$)
that, as in the lightlike case, may lead to nonunitarity problems.

\subsection{IIIc. Constitutive relations}

An analogy between the photon sector of the minimal SME
and the electrodynamics of macroscopic media has been pointed out in~\cite{Colladay}
and then extended to the case of the SME in~\cite{Kostelecky1}.
To see how this analogy works in our case,
let us introduce the displacement and magnetizing fields, respectively, as
$D_i = -\mathcal{D}_{0i}$ and $H_i = \frac12 \, \epsilon_{ijk} \mathcal{D}_{jk}$,
where $D_{\mu\nu}$ is the displacement tensor introduced in Sec. IIa.
The equations of motion, in terms of the these fields, read
\begin{equation}
\label{Eom8}
\nabla \cdot \D = 0, \;\;\; \dot{\D}= \nabla \times \HH,
\end{equation}
where $\D = (D_1,D_2,D_3)$ and $\HH = (H_1,H_2,H_3)$.
In Fourier space, $\D(t,\x) = \int \! \frac{d\omega}{2\pi} \,  \D(\omega,\x) \, e^{-i\omega t}$
and $\HH(t,\x) = \int \! \frac{d\omega}{2\pi} \,  \HH(\omega,\x) \, e^{-i\omega t}$,
we have
\begin{eqnarray}
\label{Eom13}
&& \D  = \E + g_1 \!\nabla \times \E + i\omega g_2 \B,  \\
\label{Eom14}
&& \HH = \B + g_1 \!\nabla \times \B + i\omega g_2 \E,
\end{eqnarray}
where we omitted the argument $(\omega,\x)$ for the sake of simplicity,
and we restricted ourselves to the simple case of isotropic unit external four-vector, $n^\mu = (1,0,0,0)$.
Equations~(\ref{Eom13}) and (\ref{Eom14}), which connect the displacement and magnetizing fields
to the electric and magnetic fields, are known, in the electrodynamic theory of continuous media,
as constitutive relations and completely determine the propagation properties of electromagnetic signals.

If $g_1 = 0$, the above constitutive relations describe a ``reciprocal chiral medium''  or Pasteur medium
with electric permittivity $\varepsilon = 1$, magnetic permeability $\mu = (1+\omega^2 g_2^2)^{-1}$,
and chirality parameter $\beta = g_2$. We recall that a Pasteur medium is a special case of
bi-isotropic media defined, in the Drude-Born-Fedorov representation, by
(see, e.g.,~\cite{Constitutive} and references therein)
\begin{eqnarray}
\label{Eom15}
&& \E  = \varepsilon^{-1} \D - i\omega (\beta + i\alpha) \B,  \\
\label{Eom16}
&& \HH = \mu^{-1} \B + i\omega (\beta - i\alpha) \D,
\end{eqnarray}
where $\alpha$ is the nonreciprocity parameter. Accordingly,
one could eventually gain further insight about the Myers-Pospelov, Lorentz-violating
electrodynamics by using all the known properties of the Pasteur media. Vice versa,
the Myers-Pospelov model is an example of Lagrangian description
of Pasteur media which is, to our knowledge, lacking in the literature.

At energies much lower than the Planck scale, we can recast Eqs.~(\ref{Eom13}) and (\ref{Eom14}) as
\begin{eqnarray}
\label{Eom15bis}
&& \D  = \E + i\omega (g_1 + g_2) \B,  \\
\label{Eom16bis}
&& \HH = \B - i\omega (g_1 - g_2) \E,
\end{eqnarray}
where we have neglected terms of order $\mathcal{O}(\omega^2/\mPl^2)$.
Using the equations of motion~(\ref{Eom8}), we can eliminate
the terms proportional to $g_1$ in Eqs.~(\ref{Eom15bis}) and (\ref{Eom16bis}),
which then describe a Pasteur medium
with $\varepsilon = \mu = 1$ and $\beta = g_2$.
As it is well known in the literature of bi-isotropic media~\cite{Constitutive},
linearly polarized light propagating in such a medium will experience polarization rotation
proportionally to the intensity of the chirality parameter, a result that
we have discussed, in a cosmological context, in Sec. IIc.

\section{IV. Conclusions}

Working in curved spacetime, ad using symmetry arguments, we have constructed the most general Lorentz-violating
photon Lagrangian based on dimension-five operators,
quadratic in the electromagnetic strength tensor and containing one more derivative than
the usual Maxwell term.
Assuming that Lorentz symmetry is broken by an external four-vector $n_\mu$,
we have shown that it contains only two terms, both of which break CPT symmetry.

Restricting our analysis to the case of isotropic $n_\mu$, we have then studied the generation
of cosmic magnetic fields during de Sitter inflation and cosmic birefringence in a flat Friedmann-Robertson-Walker universe.
In the former case, we have found that the creation of magnetic fields at inflation is highly suppressed at super-Hubble scales,
so they cannot explain the presence of large-scale magnetic fields detected in galaxies and galaxy clusters.
In the latter case, instead, we have found that, to the first order in the coupling parameters, the
relation between the amount of rotation of the polarization plane of linearly polarized light
and the parameters of Lorentz violation agree with the heuristic result quoted in the literature.

In Minkowski spacetime, both terms of the Lorentz-violating Lagrangian
give rise to a nontrivial dynamics, with one reducing to the well-known Myers-Pospelov Lagrangian.
Except for the case of isotropic $n_\mu$,
in which case there is violation of causality,
the full theory allows for the propagation of extra
degrees of freedom that are nonanalytic in the perturbative
coupling parameters, indicating that
nonunitary evolution may occur at the quantum level.

Finally, we have shown, in the simple case of isotropic $n_\mu$,
that the propagation of light in empty space for the Lorentz-violating theory
is equivalent to the propagation of light for the standard Maxwell theory in a continuous medium,
known as Pasteur medium.

%**********************************   Bibliography   *****************************************%


\begin{thebibliography}{99}

\bibitem{Data}            V.~A.~Kostelecky and N.~Russell,
                          %``Data Tables for Lorentz and CPT Violation,''
                          Rev.\ Mod.\ Phys.\ {\bf 83}, 11 (2011).

\bibitem{Liberati}        S.~Liberati,
                          %``Tests of Lorentz invariance: a 2013 update,''
                          Class.\ Quant.\ Grav.\ {\bf 30}, 133001 (2013).

\bibitem{Gambini}         R.~Gambini and J.~Pullin,
                          %``Nonstandard optics from quantum space-time,''
                          Phys.\ Rev.\ D {\bf 59}, 124021 (1999).

\bibitem{Kostelecky3}     V.~A.~Kostelecky and S.~Samuel,
                          %``Spontaneous Breaking of Lorentz Symmetry in String Theory,''
                          Phys.\ Rev.\ D {\bf 39}, 683 (1989).

\bibitem{Potting}         V.~A.~Kostelecky and R.~Potting,
                          %``CPT, strings, and meson factories,''
                          Phys.\ Rev.\ D {\bf 51}, 3923 (1995).

\bibitem{Kostelecky2}     V.~A.~Kostelecky,
                          %``Gravity, Lorentz violation, and the standard model,''
                          Phys.\ Rev.\ D {\bf 69}, 105009 (2004).

\bibitem{Colladay}        D.~Colladay and V.~A.~Kostelecky,
                          %``Lorentz violating extension of the standard model,''
                          Phys.\ Rev.\ D {\bf 58}, 116002 (1998).

\bibitem{Kostelecky1}     V.~A.~Kostelecky and M.~Mewes,
                          %``Electrodynamics with Lorentz-violating operators of arbitrary dimension,''
                          Phys.\ Rev.\ D {\bf 80}, 015020 (2009).

\bibitem{Kostelecky4}     A.~Kostelecky and M.~Mewes,
                          %``Neutrinos with Lorentz-violating operators of arbitrary dimension,''
                          Phys.\ Rev.\ D {\bf 85}, 096005 (2012);
                          %``Fermions with Lorentz-violating operators of arbitrary dimension,''
                          Phys.\ Rev.\ D {\bf 88}, 096006 (2013).

\bibitem{MP}              R.~C.~Myers and M.~Pospelov,
                          %``Ultraviolet modifications of dispersion relations in effective field theory,''
                          Phys.\ Rev.\ Lett.\ {\bf 90}, 211601 (2003).

\bibitem{Gubitosi}        G.~Gubitosi, L.~Pagano, G.~Amelino-Camelia, A.~Melchiorri and A.~Cooray,
                          %``A Constraint on Planck-scale Modifications to Electrodynamics with CMB polarization data,''
                          JCAP {\bf 0908}, 021 (2009).

\bibitem{Mewes}           V.~A.~Kostelecky and M.~Mewes,
                          %``Lorentz-violating electrodynamics and the cosmic microwave background,''
                          Phys.\ Rev.\ Lett.\ {\bf 99}, 011601 (2007);
                          %``Astrophysical Tests of Lorentz and CPT Violation with Photons,''
                          Astrophys.\ J.\ {\bf 689}, L1 (2008).

\bibitem{ReviewCMF}       For reviews on cosmic magnetic fields see, e.g.:
                          L.~M.~Widrow,
                          %``Origin of Galactic and Extragalactic Magnetic Fields,''
                          Rev.\ Mod.\ Phys.\ {\bf 74}, 775 (2002);
                          M.~Giovannini,
                          %``The magnetized universe,''
                          Int.\ J.\ Mod.\ Phys.\ D {\bf 13}, 391 (2004);
                          A.~Kandus, K.~E.~Kunze and C.~G.~Tsagas,
                          %``Primordial magnetogenesis,''
                          Phys.\ Rept.\ {\bf 505}, 1 (2011);
                          D.~Ryu, D.~R.~G.~Schleicher, R.~A.~Treumann, C.~G.~Tsagas and L.~M.~Widrow,
                          %``Magnetic fields in the Large-Scale Structure of the Universe,''
                          Space Sci.\ Rev.\ {\bf 166}, 1 (2012);
                          L.~M.~Widrow, D.~Ryu, D.~R.~G.~Schleicher, K.~Subramanian, C.~G.~Tsagas and R.~A.~Treumann,
                          %``The First Magnetic Fields,''
                          Space Sci.\ Rev.\ {\bf 166}, 37 (2012);
                          R.~Durrer and A.~Neronov,
                          %``Cosmological Magnetic Fields: Their Generation, Evolution and Observation,''
                          Astron.\ Astrophys.\ Rev.\ {\bf 21}, 62 (2013).

\bibitem{Turner-Widrow}   M.~S.~Turner and L.~M.~Widrow,
                          %``INFLATION PRODUCED, LARGE SCALE MAGNETIC FIELDS,''
                          Phys.\ Rev.\ D {\bf 37}, 2743 (1988);
                          B.~Ratra,
                          %``Cosmological 'seed' magnetic field from inflation,''
                          Astrophys.\ J.\ {\bf 391}, L1 (1992).

\bibitem{Banerjee}        J.~M.~Wagstaff and R.~Banerjee,
                          %``Extragalactic magnetic fields rule out electroweak phase transition magnetogenesis,''
                          arXiv:1409.4223 [astro-ph.CO].

\bibitem{Dolgov}          A.~Dolgov,
                          %``Breaking Of Conformal Invariance And Electromagnetic Field Generation In The Universe,''
                          Phys.\ Rev.\ D {\bf 48}, 2499 (1993).

\bibitem{Barrow}          J.~D.~Barrow, C.~G.~Tsagas and K.~Yamamoto,
                          %``Origin of cosmic magnetic fields: Superadiabatically amplified modes in open Friedmann universes,''
                          Phys.\ Rev.\ D {\bf 86}, 023533 (2012), and references therein.

\bibitem{Campanelli}      L.~Campanelli,
                          %``Origin of Cosmic Magnetic Fields,''
                          Phys.\ Rev.\ Lett.\ {\bf 111}, 061301 (2013).

\bibitem{Kahniashvili}    T.~Kahniashvili, A.~Brandenburg, L.~Campanelli, B.~Ratra and A.~G.~Tevzadze,
                          %``Evolution of inflation-generated magnetic field through phase transitions,''
                          Phys.\ Rev.\ D {\bf 86}, 103005 (2012);
                          L.~Campanelli,
                          %``Scaling laws in MHD turbulence,''
                          Phys.\ Rev.\ D {\bf 70}, 083009 (2004);
                          %``Evolution of Magnetic Fields in Freely Decaying Magnetohydrodynamic Turbulence,''
                          Phys.\ Rev.\ Lett.\ {\bf 98}, 251302 (2007);
                          %``Evolution of primordial magnetic fields in mean-field approximation,''
                          Eur.\ Phys.\ J.\ C {\bf 74}, 2690 (2014);
                          R.~Banerjee and K.~Jedamzik,
                          %``Are cluster magnetic fields primordial?,''
                          Phys.\ Rev.\ Lett.\ {\bf 91}, 251301 (2003);
                          %``The Evolution of cosmic magnetic fields: From the very early universe, to recombination, to the present,''
                          Phys.\ Rev.\ D {\bf 70}, 123003 (2004).

\bibitem{Campanelli2}     V.~A.~Kosteleck\'{y}, R.~Potting and S.~Samuel,
                          %``String signatures''
                          in {\it Proceedings of the 1991 Joint International Lepton-Photon
                          Symposium and Europhysics Conference on High Energy Physics},
                          %Geneva, Switzerland,
                          edited by S.~Hegarty, K.~Potter, E.~Quercigh
                          (World Scientific, Singapore, 1992);
                          O.~Bertolami and D.~F.~Mota,
                          %``Primordial magnetic fields via spontaneous breaking of Lorentz
                          %invariance,''
                          Phys.\ Lett.\ B {\bf 455}, 96 (1999);
                          L.~Campanelli, P.~Cea and G.~L.~Fogli,
                          %``Lorentz Symmetry Violation and Galactic Magnetism,''
                          Phys.\ Lett.\ B {\bf 680}, 125 (2009);
                          L.~Campanelli and P.~Cea,
                          %``Maxwell-Kosteleck\'y Electromagnetism and Cosmic Magnetization,''
                          Phys.\ Lett.\ B {\bf 675}, 155 (2009);
                          L.~Campanelli,
                          %``A Model of Universe Anisotropization,''
                          Phys.\ Rev.\ D {\bf 80}, 063006 (2009);
                          A.~P.~Kouretsis,
                          %``Cosmic magnetization in curved and Lorentz violating space-times,''
                          Eur.\ Phys.\ J.\ C {\bf 74}, 2879 (2014).

\bibitem{Birrell-Davies}  N.~D.~Birrell and P.~C.~W.~Davies,
                          {\it Quantum Fields in Curved Space}
                          (Cambridge University Press, New York, 1982).

\bibitem{Planck}          P.~A.~R.~Ade, {\it et al.}  [Planck Collaboration],
                          %``Planck 2013 results. XXII. Constraints on inflation,''
                          arXiv:1303.5082 [astro-ph.CO].

\bibitem{BICEP2}          P.~A.~R.~Ade, {\it et al.}  [BICEP2 Collaboration],
                          %``Detection of B-Mode Polarization at Degree Angular Scales by BICEP2,''
                          Phys.\ Rev.\ Lett.\ {\bf 112}, 241101 (2014).

\bibitem{Mewes2}          V.~A.~Kostelecky and M.~Mewes,
                          %``Signals for Lorentz violation in electrodynamics,''
                          Phys.\ Rev.\ D {\bf 66}, 056005 (2002).

\bibitem{Kolb-Turner}     E.~W.~Kolb and M.~S.~Turner,
                          {\it The Early Universe}
                          (Addison-Wesley, Redwood City, CA, 1990).

\bibitem{Landau}          L.~D. Landau and E.~M.~Lifshitz,
                          {\it The Quantum Mechanics: Non-Relativistic Theory}
                          (Pergamon Press, Oxford, UK, 1971).

\bibitem{Reyes}           C.~M.~Reyes,
                          %``Causality and stability for Lorentz-CPT violating electrodynamics with dimension-5 operators,''
                          Phys.\ Rev.\ D {\bf 82}, 125036 (2010).

\bibitem{Mariz}           T.~Mariz,
                          %``Radiatively induced Lorentz-violating operator of mass dimension five in QED,''
                          Phys.\ Rev.\ D {\bf 83}, 045018 (2011).

\bibitem{Mariz2}          J.~Leite and T.~Mariz,
                          %``Induced Lorentz-violating terms at finite temperature,''
                          Europhys.\ Lett.\ {\bf 99}, 21003 (2012);
                          T.~Mariz, J.~R.~Nascimento and A.~Y.~Petrov,
                          %``On the higher-derivative Lorentz-breaking terms,''
                          PoS ICMP {\bf 2012}, 022 (2012).

\bibitem{Constitutive}    S.~Ougier, I.~Chenerie, A.~Sihvola and A.~Priou,
                          %``Propagation in Bi-Isotropic Media: Effect of Different Formalisms on the Propagation Analysis,''
                          Progress In Electromagnetics Research {\bf 09}, 19 (1994).

\end{thebibliography}
\end{document}